\documentclass[aps,pra,reprint,amsmath,amssymb,longbibliography]{revtex4-2}

\usepackage{bm}
\usepackage{bbm}
\usepackage{braket}
\usepackage{mathtools}
\usepackage{amsthm}
\usepackage[colorlinks=true,citecolor=blue,linkcolor=blue,urlcolor=blue]{hyperref}

\newtheorem{proposition}{Proposition}

\DeclareMathOperator{\diag}{diag}
\DeclareMathOperator{\circmat}{circ}

\begin{document}

\title{Symmetry, circulant structure, and local unitary equivalence of  chiral $W$ states}

\author{Vladimir M. Stojanovi\'c}
\affiliation{Institut f\"{u}r Angewandte Physik, Technical
University of Darmstadt, D-64289 Darmstadt, Germany}

\date{\today}

\begin{abstract}
Endowed with projective $A_3$ permutation symmetry, chiral $W$ states are maximally entangled three-qubit states utilized in noiseless-subsystem encoding. Specifically, the chiral $W$ states $\ket{W_{+}}$ and $\ket{W_{-}}$ are the opposite-chirality, single-excitation (Hamming-weight-$1$) eigenstates of the scalar spin chirality operator. Here, their algebraic, symmetry, and entanglement properties are analyzed using general (not necessarily Hermitian) $3\times 3$ circulant matrices and the irreducible representations of the cyclic group 
$\mathbb{Z}_3 \cong A_3$. In particular, the symmetric $W$ state $\ket{W_{3}}$ and the two chiral $W$ states $\ket{W_{\pm}}$ are shown to correspond to the three irreducible representations, also forming the discrete Fourier eigenbasis of translationally invariant three-qubit operators; this is illustrated on models of both Hermitian and non-Hermitian spin rings. Furthermore, general results on $A_3$-invariant states---spanned by $\ket{W_\pm}$ and their  two-excitation counterparts $\ket{\overline{W}_\pm}$---are specialized to the chiral-$W$ manifold. It is shown that while all such $A_3$-invariant superpositions share identical bipartite and tripartite entanglement invariants, only the pure 
 basis states of Hamming weights $1$ and $2$
 are exact eigenstates of the chirality operator. Moreover, it is proved that whereas $\ket{W_{+}}$, $\ket{W_{-}}$, and $\ket{W_{3}}$ belong to the same local unitary ($\mathrm{LU}$) equivalence class under arbitrary single-qubit unitaries $U_1 \otimes U_2 \otimes U_3$, they reside in distinct equivalence classes under symmetric (global) unitaries $U^{\otimes 3}$. Finally, criteria are derived for generating $\ket{W_{\pm}}$ via $A_3$-symmetrization of product states, their continuous and discrete local stabilizer groups are characterized, and their orthogonal $S_3$-orbit is shown to realize the minimal permutation-orbit quantum code supporting a two-dimensional noiseless subsystem.
\end{abstract}

\maketitle

\section{Introduction}
Multipartite entangled states with well-defined permutation symmetries are key resources for quantum information processing and quantum metrology~\cite{Dur+:00,Acin+:00,Kraus:10PRL,Kraus:10PRA}. Similar to its Greenberger-Horne-Zeilinger (GHZ)~\cite{Song+:17,Erhard+:18,
Macri+:18} and higher-excitation Dicke~\cite{
Stojanovic+Nauth:23, 
Zhang+Jing:26,
Stojanovic+:26,Vu+:26,
Vu++:26,Nguyen+:26} counterparts, the $N$-qubit $W$ state---a balanced linear 
combination of all 
single-excitation (Hamming-weight-$1$) product states---is invariant under qubit permutation
(i.e., $S_N$-invariant, where $S_N$ denotes the permutation group). Notably, all such states with full $S_N$ symmetry reside in an $(N+1)$-dimensional subspace (commonly referred to as the symmetric sector) of the full $2^N$-dimensional Hilbert space $(\mathbbm{C}^2)^{\otimes N}$~\cite{Stojanovic+Nauth:23}. 

While most of the multipartite entangled states utilized in quantum information processing 
are fully permutationally invariant, examples of states do exist 
which lack full $S_N$
symmetry. One class of such states, defined in three-qubit systems, are
{\em chiral $W$ states}~\cite{Haase++:22}.
The chiral $W$ states 
$\ket{W_+}$ and $\ket{W_-}$
are the single-excitation (Hamming-weight-$1$) eigenstates of the normalized three-qubit scalar chirality operator with eigenvalues $+1$ and 
 $-1$, respectively. One well-established application of these states in the realm of QC pertains to implementing passive error correction 
in the form of noiseless-subsystem (NSS) qubit encoding~\cite{Zanardi+:97,
Knill+:00}.
This is a known encoding scheme for logical qubits, a three-qubit generalization of the decoherence-free subspace encoding~\cite{Lidar+:98}. While the latter is robust against global 
$Z$ dephasing, NSS encoding is insensitive to noise represented by any global Pauli operator~\cite{Knill+:00}.

Despite their growing importance in quantum-information processing protocols~\cite{Haase++:22,Reascos+:23}, several fundamental algebraic and entanglement-classification properties of $\ket{W_+}$ and $\ket{W_-}$ have not yet been systematically unified. First, while general mathematical classifications of three-qubit states invariant under permutation subgroups were recently established~\cite{Lyons+:22}, their explicit consequences for the chiral $W$ states—including the sharp distinction between arbitrary local unitary ($\mathrm{LU}$) transformations $U_1 \otimes U_2 \otimes U_3$ and uniform (global) single-qubit operations $U^{\otimes 3}$—remain to be made transparent. Second, the connection between the cyclic $\mathbb{Z}_3$ representation theory of $\ket{W_\pm}$ and the general theory of $3\times 3$ circulant matrices (including non-Hermitian matrices) provides a direct analytical framework for diagonalizing translationally invariant three-qubit Hamiltonians and quantum channels.

In this paper, a unified algebraic, group-theoretic, and entanglement characterization of the three-qubit chiral $W$ states $\ket{W_+}$ and $\ket{W_-}$ is presented. By formulating the single-excitation subspace in terms of general (not necessarily Hermitian) $3\times 3$ circulant matrices and the one-dimensional irreducible representations of the cyclic permutation group $\mathbb{Z}_3 \cong A_3$, the symmetric $W$ state $\ket{W_3}$ and the chiral states $\ket{W_\pm}$ are established as the universal discrete Fourier eigenbasis of translationally invariant three-qubit operators, as illustrated on both coherent Hermitian spin rings with chiral interactions and open non-Hermitian rings with nonreciprocal hopping. Furthermore, general results on $A_3$-invariant states~\cite{Lyons+:22} are specialized to the chiral-$W$ manifold: the strict invariance of bipartite and tripartite entanglement measures across superpositions of $\ket{W_\pm}$ and their spin-flipped two-excitation counterparts $\ket{\overline{W}_\pm}$ is proved, the dichotomy between general $\mathrm{LU}$ equivalence under $\mathrm{U}(2)^{\otimes 3}$ and inequivalence under global unitaries $U^{\otimes 3}$ is established, and explicit Bloch-angle symmetrization criteria and local stabilizer groups are derived. Finally, the orthogonal $S_3$-orbit $\{\ket{W_+}, \ket{W_-}\}$ is shown to realize a minimal permutation-orbit noiseless subsystem, and the symmetry-breaking requirements governing single-excitation subspace controllability in linear $XX$ and $XXZ$ qubit arrays are clarified.

The remainder of this paper is organized as follows. In Sec.~\ref{ChiralWintro}, the definition and basic algebraic, symmetry, and entanglement properties of the chiral $W$ states $\ket{W_+}$ and $\ket{W_-}$ are reviewed. In Sec.~\ref{sec:circulant}, the single-excitation subspace is formulated in terms of general $3\times 3$ circulant matrices and the one-dimensional irreducible representations of $\mathbb{Z}_3 \cong A_3$, establishing $\{\ket{W_3}, \ket{W_+}, \ket{W_-}\}$ as the universal discrete Fourier eigenbasis of translationally invariant operators. In Sec.~\ref{sec:examples}, this circulant-matrix framework is illustrated on two concrete three-qubit ring models: a coherent Hermitian spin ring with chiral interactions and an open non-Hermitian ring with directional hopping. In Sec.~\ref{sec:LU}, general results on $A_3$-invariant states~\cite{Lyons+:22} are specialized to the chiral-$W$ manifold, establishing the invariance of bipartite and tripartite entanglement measures, proving the dichotomy between general $\mathrm{LU}$ equivalence and global-$U^{\otimes 3}$ inequivalence, and deriving product-state symmetrization criteria and local stabilizer groups. In Sec.~\ref{sec:NSS}, the realization of a minimal permutation-orbit noiseless subsystem by $\{\ket{W_+}, \ket{W_-}\}$ and its implications for dynamical state preparation in controllable $XX$ and $XXZ$ qubit arrays are analyzed. Finally, a summary and concluding remarks are presented in Sec.~\ref{SummConcl}.

\section{Chiral $W$ states: Basic properties} \label{ChiralWintro}
The most straightforward way to introduce chiral $W$ states is to define them 
as eigenstates of the normalized three-qubit (scalar) {\em chirality} operator~\cite{Wen+:89}. 
This Hermitian operator is given by (note that units are chosen such that $\hbar=1$ throughout)
\begin{equation}\label{eq:chirality_op}
\hat{\chi} = \frac{4}{\sqrt{3}}\,\bm{S}_1 \cdot (\bm{S}_2 \times \bm{S}_3) = \frac{1}{2\sqrt{3}}\,\bm{\sigma}_1 \cdot (\bm{\sigma}_2 \times \bm{\sigma}_3),
\end{equation}
where $\bm{S}_n = \bm{\sigma}_n/2$ denotes the spin-$1/2$ operator on qubit $n$ ($n=1,2,3$).

Arguably the most interesting mathematical property of the chirality operator is 
that it conserves the total excitation number---namely, it commutes with the $z$ component of the total pseudospin operator. To determine the eigenvalues and eigenstates of the chirality operator, it is pertinent to first note that in the reordered 
three-qubit computational basis $\{|000\rangle,  |001\rangle, |010\rangle, 
|100\rangle, |110\rangle, |101\rangle, |011\rangle,
|111\rangle\}$ it adopts a block-diagonal form. It is straightforward to show that
the eigenvalues of this operator are $0$, $1$, and $-1$, with their respective degeneracies being given by 
$4$, $2$, and $2$.

\begin{figure}[t!]
\centering
\includegraphics[width=\columnwidth]{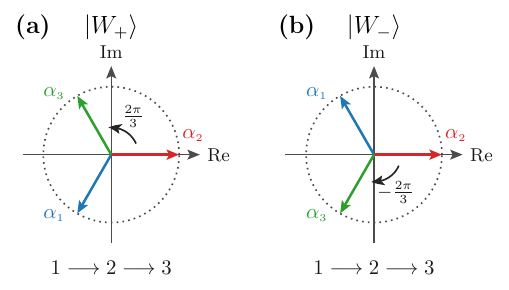}
\caption{Complex-plane phasor representation of the single-excitation coefficients $\alpha_n$ ($n=1,2,3$) for the chiral three-qubit states (a) $\ket{W_+} = (\omega^*\ket{100} + \ket{010} + \omega\ket{001})/\sqrt{3}$ and (b) $\ket{W_-} = (\omega\ket{100} + \ket{010} + \omega^*\ket{001})/\sqrt{3}$, where $\omega = e^{2\pi i/3}$. In both panels, $\alpha_1$ (blue), $\alpha_2$ (red), and $\alpha_3$ (green) lie on the circle of radius $1/\sqrt{3}$ at the cube roots of unity. Advancing cyclically across qubit sites ($1 \longrightarrow 2 \longrightarrow 3$) corresponds to a counterclockwise phase step of $2\pi/3$ in $\ket{W_+}$ and a clockwise phase step of 
$-2\pi/3$ in $\ket{W_-}$.}
\label{fig:ChiralPhasorDiagram}
\end{figure}

Among the eight eigenstates 
of the chirality operator, which form a basis of the three-qubit Hilbert space, 
there are three single-excitation ones---more precisely, three $W$-type  states, denoted by $\ket{W_+}$, $\ket{W_-}$, and $\ket{W_3}$. They satisfy the eigenvalue equations
\begin{equation}\label{eq:chirality_eig}
\hat{\chi}\ket{W_{\pm}} = \pm\ket{W_{\pm}}, \qquad \hat{\chi}\ket{W_3} = 0.
\end{equation}
The state $\ket{W_3}$, which corresponds to the eigenvalue $0$, is the symmetric (fully permutationally invariant) $W$ state
\begin{equation}
\ket{W_3} = \frac{1}{\sqrt{3}}\left(\ket{100} + \ket{010} + 
\ket{001}\right) .
\end{equation}
This state serves as the paradigmatic representative of genuine tripartite entanglement that is robust against particle loss~\cite{Dur+:00}. The two remaining single-excitation eigenstates 
of the operator $\chi$, corresponding to the eigenvalues $1$ and $-1$, respectively are the two chiral $W$ states~\cite{Haase++:22,Reascos+:23}
\begin{align}
\ket{W_{+}} &= \frac{1}{\sqrt{3}}\left(\omega^{*}\ket{100} + \ket{010} + \omega\ket{001}\right), \label{eq:Wplus} \\
\ket{W_{-}} &= \frac{1}{\sqrt{3}}\left(\omega\ket{100} + \ket{010} + \omega^{*}\ket{001}\right), \label{eq:Wminus}
\end{align}
where $\omega \equiv \exp(2\pi i/3)$ is a primitive cube root of unity. 

Unlike the uniform-phase 
$\ket{W_3}$ state, the states $\ket{W_+}$ and $\ket{W_-}$ carry opposite spatial phase windings.
The complex-plane phasor representation of the
phase winding characteristic 
of the chiral states $\ket{W_{+}}$ and  
$\ket{W_{-}}$ is depicted in 
Fig.~\ref{fig:ChiralPhasorDiagram}.

\section{Circulant-matrix and $\mathbb{Z}_3$-representation structure}\label{sec:circulant}

Having reviewed the definition and fundamental properties of chiral $W$ states in Sec.~\ref{ChiralWintro}, their underlying group-theoretic and algebraic structure within the single-excitation subspace is formalized here. Specifically, this section is organized into two parts: first, the projective $A_3$ permutation symmetry of $\ket{W_\pm}$ is analyzed in terms of the three distinct one-dimensional irreducible representations and characters of the cyclic group $\mathbb{Z}_3 \cong A_3$ (Sec.~\ref{ProjectiveA_3charZ_3}); second, general $3\times 3$ circulant matrices are employed to establish $\{\ket{W_3}, \ket{W_+}, \ket{W_-}\}$ as the universal discrete Fourier eigenbasis of translationally invariant three-qubit operators (Sec.~\ref{CircMatrDFeigen}).

\subsection{Projective $A_3$ symmetry and $\mathbb{Z}_3$ characters}
\label{ProjectiveA_3charZ_3}
Let $P_\sigma$ denote the unitary permutation operator on $(\mathbb{C}^2)^{\otimes 3}$ acting on computational basis states as
\begin{equation}
P_\sigma \ket{b_1 b_2 b_3} = \ket{b_{\sigma^{-1}(1)} b_{\sigma^{-1}(2)} b_{\sigma^{-1}(3)}}, \qquad \sigma \in S_3.
\end{equation}
The cyclic permutation $P_{(123)}$ generates the alternating subgroup $A_3 = \{e, (123), (132)\} \cong \mathbb{Z}_3$. On the single-excitation basis $\{\ket{100}, \ket{010}, \ket{001}\}$, the cyclic shift acts as $P_{(123)}\ket{100} = \ket{010}$, $P_{(123)}\ket{010} = \ket{001}$, and $P_{(123)}\ket{001} = \ket{100}$. Applying $P_{(123)}$ directly to Eqs.~\eqref{eq:Wplus} and \eqref{eq:Wminus} yields
\begin{align}
P_{(123)}\ket{W_3} &= \ket{W_3}, \label{eq:Z3_W3} \\
P_{(123)}\ket{W_+} &= \omega^* \ket{W_+}, \label{eq:Z3_Wp} \\
P_{(123)}\ket{W_-} &= \omega \ket{W_-}. \label{eq:Z3_Wm}
\end{align}
Consequently, $\ket{W_3}$, $\ket{W_+}$, and $\ket{W_-}$ carry the three distinct one-dimensional irreducible representations (characters) $\chi_0(g) = 1$, $\chi_+(g) = \omega^{-g}$, and $\chi_-(g) = \omega^{g}$ ($g \in \{0,1,2\}$) of $\mathbb{Z}_3$. At the state-vector level, $\ket{W_+}$ and $\ket{W_-}$ exhibit \emph{projective} $A_3$ permutation symmetry [transforming covariantly by global phase factors $\omega^{\mp 1}$ under $P_{(123)}$], whereas their density matrices $\rho_\pm = \ket{W_\pm}\bra{W_\pm}$ are strictly $A_3$-invariant:
\begin{equation}
P_\sigma \rho_\pm P_\sigma^\dagger = \rho_\pm, \qquad \forall \sigma \in A_3.
\end{equation}

In contrast, any odd permutation (transposition) $\tau \in S_3 \setminus A_3 = \{(12), (23), (13)\}$ reverses the orientation of the three-qubit ring and maps the two chiral sectors into each other:
\begin{align}
P_{(23)}\ket{W_+} &= \ket{W_-}, & P_{(23)}\ket{W_-} &= \ket{W_+}, \label{eq:swap23} \\
P_{(12)}\ket{W_+} &= \omega\ket{W_-}, & P_{(12)}\ket{W_-} &= \omega^*\ket{W_+}, \label{eq:swap12} \\
P_{(13)}\ket{W_+} &= \omega^*\ket{W_-}, & P_{(13)}\ket{W_-} &= \omega\ket{W_+}, \label{eq:swap13}
\end{align}
while leaving $\ket{W_3}$ invariant ($P_\tau\ket{W_3} = \ket{W_3}$). Thus, $\ket{W_3}$ spans the trivial one-dimensional irrep of $S_3$, whereas $\{\ket{W_+}, \ket{W_-}\}$ spans the two-dimensional standard irreducible representation of $S_3$.

\subsection{Circulant matrices and the discrete Fourier eigenbasis} \label{CircMatrDFeigen}
When restricted to the ordered single-excitation basis $\mathcal{B}_1 = \{\ket{100}, \ket{010}, \ket{001}\}$, the cyclic shift operator takes the fundamental $3\times 3$ circulant shift matrix form
\begin{equation}\label{eq:shift_matrix}
C \equiv P_{(123)}\big|_{\mathcal{H}_1} = \begin{pmatrix} 0 & 0 & 1 \\ 1 & 0 & 0 \\ 0 & 1 & 0 \end{pmatrix},
\end{equation}
satisfying $C^3 = I_3$ and $C^2 = C^\dagger = C^{\mathsf{T}}$. Any operator $M$ acting on $\mathcal{H}_1$ that commutes with cyclic qubit shifts ($[M, C] = 0$) is a $3\times 3$ \emph{circulant matrix}~\cite{Davis:79} generated as a matrix polynomial in $C$:
\begin{align}\label{eq:circulant_def}
M &= \circmat(c_0, c_1, c_2) \equiv c_0 I_3 + c_1 C + c_2 C^2 \nonumber \\
  &= \begin{pmatrix} c_0 & c_2 & c_1 \\ c_1 & c_0 & c_2 \\ c_2 & c_1 & c_0 \end{pmatrix}, \qquad c_0, c_1, c_2 \in \mathbb{C}.
\end{align}

Importantly, Eq.~\eqref{eq:circulant_def} holds for \emph{arbitrary complex} entries $c_0, c_1, c_2 \in \mathbb{C}$, without requiring $M$ to be Hermitian ($c_2 = c_1^*$). Every circulant matrix is normal ($[M, M^\dagger] = 0$) and is universally diagonalized by the unitary discrete Fourier transform matrix
\begin{equation}\label{eq:F3}
F_3 = \frac{1}{\sqrt{3}}\begin{pmatrix} 1 & \omega^* & \omega \\ 1 & 1 & 1 \\ 1 & \omega & \omega^* \end{pmatrix},
\end{equation}
whose columns are precisely the coordinate vectors of $\ket{W_3}$, $\ket{W_+}$, and $\ket{W_-}$ in the ordered basis $\mathcal{B}_1$. Specifically, $F_3^\dagger M F_3 = \diag(\lambda_0, \lambda_+, \lambda_-)$, where the exact eigenvalues are given by
\begin{align}
M\ket{W_3} &= (c_0 + c_1 + c_2)\ket{W_3} \equiv \lambda_0 \ket{W_3}, \label{eq:circ_eig0} \\
M\ket{W_+} &= (c_0 + c_1 \omega^* + c_2 \omega)\ket{W_+} \equiv \lambda_+ \ket{W_+}, \label{eq:circ_eigp} \\
M\ket{W_-} &= (c_0 + c_1 \omega + c_2 \omega^*)\ket{W_-} \equiv \lambda_- \ket{W_-}. \label{eq:circ_eigm}
\end{align}

Equivalently, within the group algebra $\mathbb{C}[\mathbb{Z}_3]$, the rank-one projectors onto the three basis states are the primitive orthogonal idempotents
\begin{align}
\Pi_0 &= \ket{W_3}\bra{W_3} = \frac{1}{3}\left(I_3 + C + C^2\right), \label{eq:proj0} \\
\Pi_+ &= \ket{W_+}\bra{W_+} = \frac{1}{3}\left(I_3 + \omega C + \omega^* C^2\right), \label{eq:projp} \\
\Pi_- &= \ket{W_-}\bra{W_-} = \frac{1}{3}\left(I_3 + \omega^* C + \omega C^2\right), \label{eq:projm}
\end{align}
which satisfy $\Pi_\alpha \Pi_\beta = \delta_{\alpha\beta}\Pi_\alpha$ and $\Pi_0 + \Pi_+ + \Pi_- = I_3$. Because Eqs.~\eqref{eq:circ_eig0}--\eqref{eq:circ_eigm} do not assume Hermiticity, the chiral states $\ket{W_\pm}$ constitute the universal eigenbasis not only for coherent $\mathbb{Z}_3$-symmetric Hamiltonians, but also for effective non-Hermitian Hamiltonians with nonreciprocal couplings~\cite{Metelmann+:15,Ashida+:20} and for single-excitation symmetry blocks of translationally invariant Lindbladian superoperators~\cite{Begg+:24}.

\section{Physical realizations on three-qubit rings}\label{sec:examples}
Symmetry properties of chiral $W$ states were presented using the circulant-matrix formalism and irreducible representations of the (Abelian) discrete cyclic group $\mathbb{Z}_3$ 
in Sec.~\ref{sec:circulant}.
With the aim of illustrating the utility of the circulant-matrix formulation in Eqs.~\eqref{eq:circulant_def}--\eqref{eq:circ_eigm}, two concrete physical models defined on a 
three-qubit ring are examined here: a coherent Hermitian spin ring with chiral interactions (Sec.~\ref{DMchiralRing}) and an open non-Hermitian ring with directional hopping (Sec.~\ref{NonrecHoppRing}).

\subsection{Three-qubit ring with Dzyaloshinskii--Moriya and scalar chirality interactions} \label{DMchiralRing}
Consider a translationally invariant three-qubit $XXZ$ ring. Assuming that the nearest-neighbor $XXZ$ coupling is augmented by an out-of-plane Dzyaloshinskii--Moriya (DM) interaction and a three-body scalar spin chirality term, the full (Hermitian) Hamiltonian of the ring 
is given by
\begin{align}\label{eq:H_ring}
H_{\mathrm{ring}} &= J \sum_{n=1}^{3} \left(\sigma_n^x \sigma_{n+1}^x + \sigma_n^y \sigma_{n+1}^y + \Delta \sigma_n^z \sigma_{n+1}^z\right) \nonumber \\
&\quad + D \sum_{n=1}^{3} \left(\sigma_n^x \sigma_{n+1}^y - \sigma_n^y \sigma_{n+1}^x\right) + J_\chi \hat{\chi},
\end{align}
with periodic boundary conditions ($\bm{\sigma}_4 \equiv \bm{\sigma}_1$), exchange coupling $J$, anisotropy $\Delta$, DM strength $D$, scalar chirality coupling $J_\chi$, and $\hat{\chi}$ defined in Eq.~\eqref{eq:chirality_op}.

Because $H_{\mathrm{ring}}$ conserves total magnetization $\sum_{n=1}^3 \sigma_n^z$ and commutes with cyclic shifts $P_{(123)}$, its restriction to the single-excitation basis $\mathcal{B}_1 = \{\ket{100}, \ket{010}, \ket{001}\}$ is an explicit Hermitian circulant matrix:
\begin{widetext}
\begin{equation}\label{eq:H_ring_matrix}
H_{\mathrm{ring}}\big|_{\mathcal{H}_1} = \begin{pmatrix}
-J\Delta & 2J - i\left(2D - \frac{J_\chi}{\sqrt{3}}\right) & 2J + i\left(2D - \frac{J_\chi}{\sqrt{3}}\right) \\[6pt]
2J + i\left(2D - \frac{J_\chi}{\sqrt{3}}\right) & -J\Delta & 2J - i\left(2D - \frac{J_\chi}{\sqrt{3}}\right) \\[6pt]
2J - i\left(2D - \frac{J_\chi}{\sqrt{3}}\right) & 2J + i\left(2D - \frac{J_\chi}{\sqrt{3}}\right) & -J\Delta
\end{pmatrix}.
\end{equation}
\end{widetext}
Comparing Eq.~\eqref{eq:H_ring_matrix} with Eq.~\eqref{eq:circulant_def} immediately identifies the circulant coefficients
\begin{equation}
c_0 = -J\Delta, \qquad c_1 = c_2^* = 2J + i\left(2D - \frac{J_\chi}{\sqrt{3}}\right).
\end{equation}
Substituting these coefficients into Eqs.~\eqref{eq:circ_eig0}--\eqref{eq:circ_eigm} yields the exact eigenenergies of $\ket{W_3}$, $\ket{W_+}$, and $\ket{W_-}$ by inspection:
\begin{align}
E_0 &= 4J - J\Delta, \label{eq:E0_ring} \\
E_+ &= -2J - J\Delta + 2\sqrt{3}D - J_\chi, \label{eq:Ep_ring} \\
E_- &= -2J - J\Delta - 2\sqrt{3}D + J_\chi. \label{eq:Em_ring}
\end{align}
Equations~\eqref{eq:Ep_ring} and \eqref{eq:Em_ring} show explicitly how the anti-Hermitian off-diagonal parts of $c_1 = c_2^*$ arising from the DM and scalar chirality terms lift the chiral degeneracy by $\Delta E = E_+ - E_- = 4\sqrt{3}D - 2J_\chi$. In particular, for antiferromagnetic exchange ($J > 0$) and $2\sqrt{3}D = J_\chi$, the chiral doublet $\{\ket{W_+}, \ket{W_-}\}$ forms an exact twofold-degenerate ground-state manifold separated from $\ket{W_3}$ by the energy gap $E_0 - E_\pm = 6J$.

\subsection{Non-Hermitian three-qubit ring with nonreciprocal hopping}\label{NonrecHoppRing}
In open quantum systems governed by postselected quantum trajectories or reservoir-engineered directional couplings~\cite{Metelmann+:15,Ashida+:20,Begg+:24}, the effective single-excitation dynamics is described by a non-Hermitian tight-binding ring with asymmetric (nonreciprocal) hopping amplitudes $J_R \neq J_L$:
\begin{equation}\label{eq:H_nH}
H_{\mathrm{nH}} = \sum_{n=1}^{3} \left(\epsilon_0 \ket{1_n}\bra{1_n} + J_R \ket{1_{n+1}}\bra{1_n} + J_L \ket{1_n}\bra{1_{n+1}}\right),
\end{equation}
where $\ket{1_n}$ denotes the single-excitation state on site $n$ (with $\ket{1_4} \equiv \ket{1_1}$) and $\epsilon_0 \in \mathbb{C}$ includes uniform on-site detuning and loss. In the ordered basis $\mathcal{B}_1 = \{\ket{100}, \ket{010}, \ket{001}\}$, $H_{\mathrm{nH}}$ takes the non-Hermitian circulant matrix form
\begin{equation}\label{eq:H_nH_matrix}
H_{\mathrm{nH}}\big|_{\mathcal{H}_1} = \begin{pmatrix}
\epsilon_0 & J_L & J_R \\
J_R & \epsilon_0 & J_L \\
J_L & J_R & \epsilon_0
\end{pmatrix} = \circmat(\epsilon_0, J_R, J_L).
\end{equation}

Applying Eqs.~\eqref{eq:circ_eig0}--\eqref{eq:circ_eigm} directly yields the complex chiral spectrum of $H_{\mathrm{nH}}$:
\begin{align}
\lambda_0 &= \epsilon_0 + J_R + J_L, \label{eq:lam0_nH} \\
\lambda_+ &= \epsilon_0 - \frac{1}{2}(J_R + J_L) - i\frac{\sqrt{3}}{2}(J_R - J_L), \label{eq:lamp_nH} \\
\lambda_- &= \epsilon_0 - \frac{1}{2}(J_R + J_L) + i\frac{\sqrt{3}}{2}(J_R - J_L). \label{eq:lamm_nH}
\end{align}
Whereas the Hermitian DM and chirality couplings in Eqs.~\eqref{eq:Ep_ring}--\eqref{eq:Em_ring} split the \emph{real} energies of $\ket{W_+}$ and $\ket{W_-}$, nonreciprocal hopping ($J_R \neq J_L$) splits their \emph{imaginary} parts $\mathrm{Im}(\lambda_\pm) = \mp \frac{\sqrt{3}}{2}(J_R - J_L)$, inducing selective chiral amplification or decay of $\ket{W_+}$ relative to $\ket{W_-}$ without mixing the eigenstates.

\section{$\mathrm{LU}$ equivalence and $A_3$-invariant state properties}\label{sec:LU}

Having established the circulant-matrix formulation and physical ring realizations of chiral $W$ states in Secs.~\ref{sec:circulant} and~\ref{sec:examples}, $\ket{W_\pm}$ are situated here within the broader framework of $A_3$-invariant three-qubit states~\cite{Lyons+:22}, and their local unitary ($\mathrm{LU}$) transformation properties are examined. Specifically, this section is organized as follows: first, general $A_3$-invariant superpositions of $\ket{W_\pm}$ and their two-excitation counterparts $\ket{\overline{W}_\pm}$ are analyzed with respect to scalar spin chirality (Sec.~\ref{A_3invScalChirality}), and the invariance of their bipartite and tripartite entanglement measures across this manifold is proved (Sec.~\ref{InvBipartTripart}); next, the sharp dichotomy between general $\mathrm{LU}$ equivalence under $\mathrm{U}(2)^{\otimes 3}$ and inequivalence under symmetric global unitaries $U^{\otimes 3}$ is established (Sec.~\ref{LUequivInequiv}); and finally, explicit product-state $A_3$-symmetrization criteria are derived (Sec.~\ref{ProdStateA_3Symm}) and the continuous and discrete local stabilizer groups of $\ket{W_\pm}$ are characterized (Sec.~\ref{ContDiscrStabGroups}).

\subsection{General $A_3$-invariant three-qubit states and scalar chirality}
\label{A_3invScalChirality}
It was previously demonstrated (although using a different notation and without direct reference to chiral $W$ states)~\cite{Lyons+:22} that any $A_3$-invariant but not $S_3$-invariant three-qubit state can be written---up to a global phase factor---as a normalized linear combination of the state 
$\ket{W_+}$ and its two-excitation (Hamming-weight-$2$) counterpart obtained through local change of basis $|0\rangle \rightleftarrows|1\rangle$ (a generalized two-excitation Dicke state) 
\begin{equation}
\ket{\overline{W}_+}=\frac{1}{\sqrt{3}}\left(\omega^{*}
\ket{011} + \ket{101} + \omega\ket{110}\right),\label{eq:Wbarplus}
\end{equation}
or as a complex conjugate 
(in the computational basis) of such a linear combination. In other words, any such state can be written in the form
\begin{equation}
\ket{\psi}=\mu_1 
\ket{W_+} +\mu_2\ket{\overline{W}_+} ,
\label{eq:A3invStates}
\end{equation}
where $\mu_1$ and $\mu_2$ are complex numbers that satisfy the normalization condition $|\mu_1|^2+|\mu_2|^2 = 1$, or as the complex conjugate of this linear combination. Note that the complex conjugate of the state $\ket{W_+}$ is $\ket{W_-}$, while the conjugate of $\ket{\overline{W}_+}$ is the analogously defined state $\ket{\overline{W}_{-}}$.

In this parametrization, the single-excitation chiral state $\ket{W_+}$ and its complex conjugate $\ket{W_-}$ correspond to the special case $\mu_1 = 1$ and $\mu_2 = 0$, whereas the two-excitation states $\ket{\overline{W}_+}$ and $\ket{\overline{W}_-}$ correspond to $\mu_1 = 0$ and $\mu_2 = 1$. Under the global spin-flip operation $X^{\otimes 3}$, the normalized scalar chirality operator $\hat{\chi}$ satisfies the symmetry relation
\begin{equation}
X^{\otimes 3}\hat{\chi}X^{\otimes 3} = -\hat{\chi},
\label{eq:SpinFlipChirality}
\end{equation}
which implies that $\ket{\overline{W}_+}$ and $\ket{\overline{W}_-}$ are also exact eigenstates of $\hat{\chi}$, but with reversed eigenvalues $-1$ and $+1$, respectively. Because $\ket{W_+}$ and $\ket{\overline{W}_+}$ belong to orthogonal eigenvalue sectors of $\hat{\chi}$, the scalar chirality expectation value of an arbitrary normalized superposition $\ket{\psi} = \mu_1\ket{W_+} + \mu_2\ket{\overline{W}_+}$ in Eq.~\eqref{eq:A3invStates} (and likewise of its complex conjugate $\ket{\psi^*}$) takes the simple form
\begin{equation}
\braket{\psi|\hat{\chi}|\psi} = |\mu_1|^2 - |\mu_2|^2,
\label{eq:ChiralityExpectation}
\end{equation}
while $\braket{\psi^*|\hat{\chi}|\psi^*} = |\mu_2|^2 - |\mu_1|^2$. Thus, although every state in Eq.~\eqref{eq:A3invStates} shares the same projective $A_3$ permutation symmetry, only the pure Hamming-weight-$1$ and Hamming-weight-$2$ basis states ($\mu_1\mu_2 = 0$) are exact eigenstates of $\hat{\chi}$ attaining the extremal chiralities $\pm 1$; generic superpositions with $\mu_1\mu_2 \neq 0$ mix opposite-chirality subspaces and exhibit reduced chirality magnitude $|\braket{\psi|\hat{\chi}|\psi}| < 1$.

\subsection{Invariance of bipartite and tripartite entanglement measures} \label{InvBipartTripart}
In sharp contrast to the chirality expectation value, all bipartite and tripartite entanglement measures remain strictly invariant across the entire $(\mu_1,\mu_2)$ manifold, as required by the fact that all states in Eq.~\eqref{eq:A3invStates} belong to the same $\mathrm{LU}$ equivalence class~\cite{Lyons+:22}. To make this manifest, consider first the single-qubit reduced density matrices $\rho_n = \mathrm{Tr}_{\bar{n}}(\ket{\psi}\bra{\psi})$ ($n=1,2,3$), obtained by tracing out the remaining two qubits $\bar{n}$. Owing to the cyclic $A_3$ invariance of $\ket{\psi}$, all three single-qubit marginals are identical and read explicitly in the computational basis $\{\ket{0},\ket{1}\}$ as
\begin{equation}
\rho_1 = \rho_2 = \rho_3 = \frac{1}{3}
\begin{pmatrix}
1 + |\mu_1|^2 & -\mu_1\mu_2^* \\
-\mu_1^*\mu_2 & 1 + |\mu_2|^2
\end{pmatrix}.
\label{eq:SingleQubitRDM}
\end{equation}
Using the normalization condition $|\mu_1|^2 + |\mu_2|^2 = 1$, the characteristic polynomial of $\rho_n$ simplifies to $\lambda^2 - \lambda + 2/9 = 0$, yielding the constant Schmidt eigenvalues $\lambda_{\max} = 2/3$ and $\lambda_{\min} = 1/3$ independent of $\mu_1$ and $\mu_2$. Consequently, both the single-qubit purity $\mathcal{P}_n \equiv \mathrm{Tr}(\rho_n^2)$ and the bipartite linear entropy $S_L^{(n)} \equiv 2[1 - \mathrm{Tr}(\rho_n^2)]$ across every $1\vert 2$ bipartition take the constant values
\begin{equation}
\mathcal{P}_n = \frac{5}{9}, \qquad S_L^{(n)} = \frac{8}{9} \quad (n=1,2,3).
\label{eq:PurityLinearEntropy}
\end{equation}

The pairwise and genuine tripartite entanglement properties of Eq.~\eqref{eq:A3invStates} are likewise completely independent of $(\mu_1,\mu_2)$. For any two-qubit reduced density matrix $\rho_{ij} = \mathrm{Tr}_k(\ket{\psi}\bra{\psi})$ ($i < j$, with $k$ labeling the traced-out qubit), Wootters' pairwise concurrence is defined as $C_{ij} = \max\{0, \sqrt{\nu_1} - \sqrt{\nu_2} - \sqrt{\nu_3} - \sqrt{\nu_4}\}$, where $\nu_1 \ge \nu_2 \ge \nu_3 \ge \nu_4$ are the eigenvalues of the spin-flipped matrix $\rho_{ij}(Y\otimes Y)\rho_{ij}^*(Y\otimes Y)$~\cite{Wootters:98}. Evaluating this expression on Eq.~\eqref{eq:A3invStates} yields equal pairwise concurrences across all three qubit pairs:
\begin{equation}
C_{12} = C_{23} = C_{13} = \frac{2}{3}.
\label{eq:ConcurrenceA3}
\end{equation}
Finally, genuine tripartite entanglement is quantified by the Coffman--Kundu--Wootters three-tangle $\tau_3 = C_{1(23)}^2 - C_{12}^2 - C_{13}^2$, where $C_{1(23)} = 2\sqrt{\det\rho_1} = \sqrt{2[1 - \mathrm{Tr}(\rho_1^2)]}$ denotes the bipartite concurrence between qubit $1$ and the remaining pair $(23)$~\cite{Coffman+:00}. Substituting $C_{1(23)}^2 = S_L^{(1)} = 8/9$ and $C_{12}^2 = C_{13}^2 = 4/9$ gives
\begin{equation}
\tau_3 = \frac{8}{9} - \frac{4}{9} - \frac{4}{9} = 0
\label{eq:ThreeTangleA3}
\end{equation}
identically for all $\mu_1$ and $\mu_2$. Every state in Eq.~\eqref{eq:A3invStates} therefore resides strictly within the $W$ entanglement class~\cite{Dur+:00}, carrying maximal pairwise entanglement for a symmetric three-qubit state while remaining sharply distinguished from generic superpositions by its extremal scalar chirality when $\mu_1\mu_2 = 0$.

\subsection{General $\mathrm{LU}$ equivalence versus global-$U^{\otimes 3}$ inequivalence}
\label{LUequivInequiv}
It is also instructive to examine chiral $W$ states in the context of local unitary ($\mathrm{LU}$) equivalence. Recall that two pure states are $\mathrm{LU}$ equivalent if one can be transformed into the other by applying a unitary operation qubitwise, thereby partitioning the state space into equivalence classes under local basis changes. Specifically, three-qubit states $\ket{\psi}$ and $\ket{\phi}$ are $\mathrm{LU}$ equivalent if there exist single-qubit unitaries $U_1, U_2, U_3 \in \mathrm{U}(2)$ such that
\begin{equation}\label{eq:LU_def}
\ket{\phi} = (U_1 \otimes U_2 \otimes U_3)\ket{\psi},
\end{equation}
where $U_n$ acts on qubit $n$. Such $\mathrm{LU}$ equivalence implies that the two states share identical entanglement properties---including the same Schmidt spectra (and hence equal bipartite entanglement across every bipartition), three-tangle, and all polynomial $\mathrm{LU}$ invariants~\cite{Acin+:00,Kraus:10PRL,Kraus:10PRA}---consistent with the invariance of the entanglement measures established in Eqs.~\eqref{eq:SingleQubitRDM}--\eqref{eq:ThreeTangleA3}.

While Theorem~1 of Ref.~\cite{Lyons+:22} proves via a sequence of non-diagonal $\mathrm{SU}(2)$ weight-mixing rotations that every $A_3$-invariant state of the form $\mu_1\ket{W_+} + \mu_2\ket{\overline{W}_+}$ [Eq.~\eqref{eq:A3invStates}] is $\mathrm{LU}$ equivalent to $\ket{W_+}$, restricting to the single-excitation chiral manifold simplifies this equivalence dramatically while revealing a sharp dichotomy between arbitrary local unitaries $\mathrm{U}(2)^{\otimes 3}$ and symmetric (global) single-qubit unitaries $U^{\otimes 3}$:

\begin{proposition}[$\mathrm{LU}$ equivalence and global-$U^{\otimes 3}$ inequivalence]\label{prop:LU}
Let $V = \diag(1, \omega) \in \mathrm{U}(2)$ denote the single-qubit $\mathbb{Z}_3$ phase gate (with $V^\dagger = \diag(1, \omega^*)$). Then:
\begin{enumerate}
\item[\textup{(i)}] The states $\ket{W_+}$, $\ket{W_-}$, and $\ket{W_3}$ all belong to the \emph{same} $\mathrm{LU}$ equivalence class under $\mathrm{U}(2)^{\otimes 3}$, being explicitly interconverted by diagonal local phase gates:
\begin{align}
\ket{W_3} &= (V \otimes I \otimes V^\dagger)\ket{W_+}, \label{eq:Wp_to_W3} \\
\ket{W_-} &= (V^\dagger \otimes I \otimes V)\ket{W_+}. \label{eq:Wp_to_Wm}
\end{align}
\item[\textup{(ii)}] Conversely, $\ket{W_+}$, $\ket{W_-}$, and $\ket{W_3}$ are pairwise \emph{inequivalent} under symmetric (global) single-qubit unitaries of the form $U^{\otimes 3}$ ($U \in \mathrm{U}(2)$); that is, no $U \in \mathrm{U}(2)$ and phase $\theta \in \mathbb{R}$ exist such that $U^{\otimes 3}\ket{W_a} = e^{i\theta}\ket{W_b}$ for $a \neq b \in \{3, +, -\}$.
\end{enumerate}
\end{proposition}

\begin{proof}
For part (i), direct action of $V = \diag(1,\omega)$ on single-qubit computational basis states gives $V\ket{0} = \ket{0}$ and $V\ket{1} = \omega\ket{1}$. Applying $V \otimes I \otimes V^\dagger$ to $\ket{W_+}$ yields
\begin{align}
\lefteqn{(V \otimes I \otimes V^\dagger)\ket{W_+}} \nonumber \\
&\quad = \frac{1}{\sqrt{3}}\left(\omega \omega^*\ket{100} + \ket{010} + \omega^* \omega\ket{001}\right) \nonumber \\
&\quad = \frac{1}{\sqrt{3}}\left(\ket{100} + \ket{010} + \ket{001}\right) = \ket{W_3},
\end{align}
since $\omega\omega^* = 1$. Similarly, using $\omega^2 = \omega^*$ and $(\omega^*)^2 = \omega$, we obtain
\begin{align}
\lefteqn{(V^\dagger \otimes I \otimes V)\ket{W_+}} \nonumber \\
&\quad = \frac{1}{\sqrt{3}}\left((\omega^*)^2\ket{100} + \ket{010} + \omega^2\ket{001}\right) \nonumber \\
&\quad = \frac{1}{\sqrt{3}}\left(\omega\ket{100} + \ket{010} + \omega^*\ket{001}\right) = \ket{W_-}.
\end{align}

For part (ii), note that every uniform tensor-product unitary $U^{\otimes 3}$ commutes with all qubit permutations: $[U^{\otimes 3}, P_\sigma] = 0$ for all $\sigma \in S_3$. Suppose $U^{\otimes 3}\ket{W_+} = e^{i\theta}\ket{W_-}$ for some $U \in \mathrm{U}(2)$ and $\theta \in \mathbb{R}$. Acting on both sides with $P_{(123)}$ and using Eqs.~\eqref{eq:Z3_Wp}--\eqref{eq:Z3_Wm} gives
\begin{align}
e^{i\theta}\omega\ket{W_-}
&= P_{(123)}\left(e^{i\theta}\ket{W_-}\right) = P_{(123)} U^{\otimes 3}\ket{W_+} \nonumber \\
&= U^{\otimes 3} P_{(123)}\ket{W_+} = \omega^* U^{\otimes 3}\ket{W_+} \nonumber \\
&= e^{i\theta}\omega^*\ket{W_-},
\end{align}
which implies $\omega = \omega^*$, a contradiction. The same character-conservation argument rules out $U^{\otimes 3}$-equivalence between $\ket{W_3}$ (with $\mathbb{Z}_3$ character $\chi_0 = 1$) and $\ket{W_\pm}$ (with characters $\chi_\pm = \omega^{\mp 1}$).
\end{proof}

Proposition~\ref{prop:LU} clarifies why $\ket{W_+}$, $\ket{W_-}$, and $\ket{W_3}$ share identical single-qubit reduced density matrices $\rho_n = \frac{2}{3}\ket{0}\bra{0} + \frac{1}{3}\ket{1}\bra{1}$ [cf.\ Eq.~\eqref{eq:SingleQubitRDM} for $\mu_1=1,\mu_2=0$] and identical polynomial entanglement invariants, yet cannot be interconverted by uniform global fields $U^{\otimes 3}$: switching chirality requires either spatially inhomogeneous local phase control ($V \otimes I \otimes V^\dagger$) or a nonlocal two-qubit transposition ($P_{(23)}$).

\subsection{Product-state $A_3$-symmetrization criteria for chiral $W$ states} \label{ProdStateA_3Symm}
In Ref.~\cite{Lyons+:22} (Proposition~9), it was shown that $A_3$-symmetrizing a three-qubit product state
\begin{equation}
\ket{\Phi} = \ket{\phi_1}\otimes\ket{\phi_2}\otimes\ket{\phi_3}
\end{equation}
with a nontrivial character yields a nonzero $A_3$-invariant (non-$S_3$-invariant) state if and only if the single-qubit states $\ket{\phi_1}$, $\ket{\phi_2}$, and $\ket{\phi_3}$ are not all mutually collinear. To isolate the single-excitation chiral states $\ket{W_\pm}$ without admixture from the two-excitation sector $\mathcal{H}_2$ (spanned by $\ket{\overline{W}_\pm}$), we extend the group-algebra projectors $\Pi_\pm$ to the full three-qubit Hilbert space $(\mathbb{C}^2)^{\otimes 3}$ as
\begin{equation}\label{eq:FullProj}
\tilde{\Pi}_\pm = \frac{1}{3}\left(I + \omega^{\pm 1} P_{(123)} + \omega^{\mp 1} P_{(132)}\right),
\end{equation}
and parameterize each single-qubit state up to an overall phase as
\begin{equation}\label{eq:BlochParam}
\ket{\phi_j} = \cos\left(\frac{\theta_j}{2}\right)\ket{0} + e^{i\varphi_j}\sin\left(\frac{\theta_j}{2}\right)\ket{1} \qquad (j=1,2,3),
\end{equation}
where $\theta_j \in [0,\pi]$ and $\varphi_j \in [0,2\pi)$.

\begin{proposition}[Pure chiral-$W$ projection from product states]\label{prop:prod_sym}
Let $\ket{\Phi} = \ket{\phi_1}\otimes\ket{\phi_2}\otimes\ket{\phi_3}$ be a normalized three-qubit product state parameterized by Eq.~\eqref{eq:BlochParam}. Then $\tilde{\Pi}_\pm\ket{\Phi}$ is a nonzero multiple of $\ket{W_\pm}$ if and only if:
\begin{enumerate}
\item[\textup{(i)}] the single-excitation chiral overlap is nonvanishing,
\begin{align}\label{eq:ProdCond1}
&\omega^{\pm 1}\sin\frac{\theta_1}{2}\cos\frac{\theta_2}{2}\cos\frac{\theta_3}{2}e^{i\varphi_1} \nonumber \\
&\quad + \cos\frac{\theta_1}{2}\sin\frac{\theta_2}{2}\cos\frac{\theta_3}{2}e^{i\varphi_2} \nonumber \\
&\quad + \omega^{\mp 1}\cos\frac{\theta_1}{2}\cos\frac{\theta_2}{2}\sin\frac{\theta_3}{2}e^{i\varphi_3} \neq 0,
\end{align}
and
\item[\textup{(ii)}] the two-excitation chiral overlap vanishes identically,
\begin{align}\label{eq:ProdCond2}
&\omega^{\pm 1}\cos\frac{\theta_1}{2}\sin\frac{\theta_2}{2}\sin\frac{\theta_3}{2}e^{i(\varphi_2+\varphi_3)} \nonumber \\
&\quad + \sin\frac{\theta_1}{2}\cos\frac{\theta_2}{2}\sin\frac{\theta_3}{2}e^{i(\varphi_1+\varphi_3)} \nonumber \\
&\quad + \omega^{\mp 1}\sin\frac{\theta_1}{2}\sin\frac{\theta_2}{2}\cos\frac{\theta_3}{2}e^{i(\varphi_1+\varphi_2)} = 0.
\end{align}
\end{enumerate}
\end{proposition}

\begin{proof}
Because $\tilde{\Pi}_\pm$ commutes with the total excitation-number operator $\hat{N} = \sum_{j=1}^3 \ket{1}_j\!\bra{1}$, it preserves the Hamming-weight decomposition
\begin{equation}
(\mathbb{C}^2)^{\otimes 3} = \mathcal{H}_0 \oplus \mathcal{H}_1 \oplus \mathcal{H}_2 \oplus \mathcal{H}_3.
\end{equation}
Since $\ket{000} \in \mathcal{H}_0$ and $\ket{111} \in \mathcal{H}_3$ are totally symmetric under $S_3$, they carry the trivial representation of $\mathbb{Z}_3$ and are annihilated by $\tilde{\Pi}_\pm$:
\begin{equation}
\tilde{\Pi}_\pm\ket{000} = \tilde{\Pi}_\pm\ket{111} = 0.
\end{equation}
Consequently, $\tilde{\Pi}_\pm\ket{\Phi}$ projects strictly onto the direct sum of the single-excitation and two-excitation chiral subspaces,
\begin{equation}
\tilde{\Pi}_\pm\ket{\Phi} = \braket{W_\pm|\Phi}\ket{W_\pm} + \braket{\overline{W}_\pm|\Phi}\ket{\overline{W}_\pm},
\end{equation}
where $\ket{\overline{W}_\pm}$ are the spin-flipped two-excitation chiral states defined in Eq.~\eqref{eq:Wbarplus}. Expanding $\ket{\Phi} = \bigotimes_{j=1}^3\ket{\phi_j}$ in the computational basis and taking inner products with $\ket{W_\pm}$ and $\ket{\overline{W}_\pm}$ yields
\begin{align}
\braket{W_\pm|\Phi} &= \frac{1}{\sqrt{3}}\Bigl(\omega^{\pm 1}\sin\frac{\theta_1}{2}\cos\frac{\theta_2}{2}\cos\frac{\theta_3}{2}e^{i\varphi_1} \nonumber \\
&\quad + \cos\frac{\theta_1}{2}\sin\frac{\theta_2}{2}\cos\frac{\theta_3}{2}e^{i\varphi_2} \nonumber \\
&\quad + \omega^{\mp 1}\cos\frac{\theta_1}{2}\cos\frac{\theta_2}{2}\sin\frac{\theta_3}{2}e^{i\varphi_3}\Bigr)
\end{align}
and
\begin{align}
\braket{\overline{W}_\pm|\Phi} &= \frac{1}{\sqrt{3}}\Bigl(\omega^{\pm 1}\cos\frac{\theta_1}{2}\sin\frac{\theta_2}{2}\sin\frac{\theta_3}{2}e^{i(\varphi_2+\varphi_3)} \nonumber \\
&\quad + \sin\frac{\theta_1}{2}\cos\frac{\theta_2}{2}\sin\frac{\theta_3}{2}e^{i(\varphi_1+\varphi_3)} \nonumber \\
&\quad + \omega^{\mp 1}\sin\frac{\theta_1}{2}\sin\frac{\theta_2}{2}\cos\frac{\theta_3}{2}e^{i(\varphi_1+\varphi_2)}\Bigr).
\end{align}
Requiring $\braket{W_\pm|\Phi} \neq 0$ and $\braket{\overline{W}_\pm|\Phi} = 0$ immediately yields conditions~\textup{(i)} and~\textup{(ii)}.
\end{proof}

As a concrete illustration of Proposition~\ref{prop:prod_sym}, taking the localized single-excitation state $\ket{\Phi} = \ket{100}$ yields
\begin{equation}
\tilde{\Pi}_\pm\ket{100} = \frac{\omega^{\pm 1}}{\sqrt{3}}\ket{W_\pm}
\end{equation}
directly. More generally, any tilted product state with a single excited qubit,
\begin{equation}
\ket{\Phi} = \ket{\phi_1}\otimes\ket{0}\otimes\ket{0} \qquad (\ket{\phi_1} \neq \ket{0}),
\end{equation}
satisfies condition~\textup{(ii)} identically and projects cleanly onto the single-excitation chiral manifold,
\begin{equation}
\tilde{\Pi}_\pm\left(\ket{\phi_1}\otimes\ket{0}\otimes\ket{0}\right) = \frac{\omega^{\pm 1}}{\sqrt{3}}\braket{1|\phi_1}\ket{W_\pm},
\end{equation}
without populating the two-excitation sector spanned by $\ket{\overline{W}_\pm}$.

\subsection{Continuous and discrete local stabilizer groups}
\label{ContDiscrStabGroups}

Proposition~15 of Ref.~\cite{Lyons+:22} establishes that the connected identity component of the local unitary stabilizer group of any $A_3$-invariant generalized Dicke state of Hamming weight $w \neq n/2$ consists solely of collective $Z$-rotations. For the chiral $W$ states ($n=3$, $w=1$), we can characterize both the continuous stabilizer subgroup and its discrete $\mathbb{Z}_3$ ray-stabilizer extension explicitly:

\begin{proposition}[Stabilizer structure of $\ket{W_\pm}$]\label{prop:stab}
Let the projective local unitary stabilizer group of $\ket{W_\pm}$ be defined by
\begin{align}\label{eq:stab_ray_def}
\mathrm{Stab}_{\mathrm{ray}}(\ket{W_\pm}) = \Bigl\{ &(e^{i\alpha}, U_1, U_2, U_3) \in \mathrm{U}(1) \times \mathrm{U}(2)^{\otimes 3} : \nonumber \\
&e^{i\alpha}(U_1 \otimes U_2 \otimes U_3)\ket{W_\pm} = \ket{W_\pm} \Bigr\}.
\end{align}
Then:
\begin{enumerate}
\item[\textup{(i)}] The connected identity component of $\mathrm{Stab}_{\mathrm{ray}}(\ket{W_\pm})$ is the one-parameter collective $Z$-rotation group
\begin{equation}\label{eq:stab0}
\mathrm{Stab}_0(\ket{W_\pm}) = \left\{ \left(e^{-it}, R_z(t)^{\otimes 3}\right) : t \in \mathbb{R} \right\},
\end{equation}
where $R_z(t) = \diag(1, e^{it})$.
\item[\textup{(ii)}] Among global single-qubit operations $U^{\otimes 3}$, the only unitaries stabilizing the ray $\ket{W_\pm}$ are the diagonal phase rotations $R_z(t)^{\otimes 3}$ in Eq.~\eqref{eq:stab0} (up to an overall phase).
\item[\textup{(iii)}] Combining local phase gates with the cyclic permutation $P_{(123)}$ enlarges the ray stabilizer by a discrete $\mathbb{Z}_3$ factor generated by $(\omega^{\pm 1}, P_{(123)})$, reflecting the projective $\mathbb{Z}_3$ invariance
\begin{equation}\label{eq:proj_Z3_stab}
P_{(123)}\ket{W_\pm} = \omega^{\mp 1}\ket{W_\pm}.
\end{equation}
\end{enumerate}
\end{proposition}

\begin{proof}
By Proposition~\ref{prop:LU}(i), the chiral states are related to the symmetric state $\ket{W_3}$ via
\begin{equation}
\ket{W_\pm} = D_\pm \ket{W_3},
\end{equation}
where $D_+ = V^\dagger \otimes I \otimes V$ and $D_- = V \otimes I \otimes V^\dagger$ are diagonal local phase gates. Conjugating the connected stabilizer subgroup $\mathrm{Stab}_0(\ket{W_3})$~\cite{Lyons+:22} by $D_\pm$ leaves every diagonal gate $R_z(t)^{\otimes 3}$ invariant because
\begin{equation}
\left[D_\pm, R_z(t)^{\otimes 3}\right] = 0 \qquad \forall\, t \in \mathbb{R},
\end{equation}
which immediately establishes Eq.~\eqref{eq:stab0}.

For part~\textup{(ii)}, suppose that $U^{\otimes 3}\ket{W_\pm} = e^{i\alpha}\ket{W_\pm}$ for some $U \in \mathrm{U}(2)$ and $\alpha \in \mathbb{R}$. Conjugating both sides by $D_\pm^\dagger$ yields
\begin{equation}\label{eq:conj_stab}
\left(D_\pm^\dagger U^{\otimes 3} D_\pm\right)\ket{W_3} = e^{i\alpha}\ket{W_3}.
\end{equation}
Because the single-qubit reduced density matrices of $\ket{W_3}$ take the nondegenerate diagonal form
\begin{equation}
\rho_n = \diag\!\left(\frac{2}{3}, \frac{1}{3}\right) \qquad (n=1,2,3),
\end{equation}
every local unitary stabilizing $\ket{W_3}$ must commute with $\rho_n$ and hence be diagonal in the computational basis. Consequently, $U = e^{i\gamma}R_z(t)$ for some $\gamma, t \in \mathbb{R}$, which indeed satisfies
\begin{equation}
U^{\otimes 3}\ket{W_\pm} = e^{i(3\gamma+t)}\ket{W_\pm}.
\end{equation}
Finally, part~\textup{(iii)} follows directly from Eqs.~\eqref{eq:Z3_Wp}--\eqref{eq:Z3_Wm}.
\end{proof}

\section{Noiseless-subsystem encoding and subspace controllability}\label{sec:NSS}

Building on the algebraic and symmetry structure established in Secs.~\ref{sec:circulant}--\ref{sec:LU}, two direct quantum-information applications of chiral $W$ states are examined here: first, the orthogonal $S_3$-orbit $\{\ket{W_+}, \ket{W_-}\}$ is shown to realize the minimal permutation-orbit quantum code supporting a two-dimensional noiseless subsystem (Sec.~\ref{MinPermOrbitNS}); second, the implications of these symmetry and $\mathrm{LU}$ properties for dynamical state preparation in controllable linear $XX$ and $XXZ$ qubit arrays are analyzed (Sec.~\ref{ControlImplic}).

\subsection{Minimal permutation-orbit quantum code and noiseless subsystem} \label{MinPermOrbitNS}
In Sec.~V\,A of Ref.~\cite{Lyons+:22}, a general framework was proposed for constructing quantum error-correcting codes from orthogonal states lying within the $S_n$-orbit $\{P_\sigma\ket{\psi} : \sigma \in S_n\}$ of a $G$-invariant state $\ket{\psi}$, leaving the detailed properties of such permutation-orbit codes to future study. Here, we observe that the chiral $W$ pair $\{\ket{W_+}, \ket{W_-}\}$ provides the minimal ($n=3$) exact realization of such a permutation-orbit code, and that this structure is directly responsible for its utility in noiseless-subsystem (NS) qubit encoding~\cite{Knill+:00,Zanardi+:97,Haase+:22}:

\begin{proposition}[Minimal $S_3$-orbit noiseless subsystem]\label{prop:NSS}
The projective $S_3$-orbit of $\ket{W_+}$ consists exclusively of the two orthogonal states $\{\ket{W_+}, \ket{W_-}\}$. Moreover, the two-dimensional subspace
\begin{equation}\label{eq:CNS_def}
\mathcal{C}_{\mathrm{NS}} = \mathrm{span}\{\ket{W_+}, \ket{W_-}\}
\end{equation}
is invariant under all global single-qubit unitaries $U^{\otimes 3}$ ($U \in \mathrm{U}(2)$) and forms a noiseless subsystem encoding one logical qubit protected against arbitrary collective noise channels of the form
\begin{equation}\label{eq:coll_noise}
\mathcal{E}(\rho) = \sum_\mu E_\mu^{\otimes 3}\rho (E_\mu^\dagger)^{\otimes 3}.
\end{equation}
\end{proposition}

\begin{proof}
By Eqs.~\eqref{eq:Z3_Wp}--\eqref{eq:swap13}, every even permutation $\sigma \in A_3$ maps $\ket{W_+}$ to itself up to a phase $\omega^{\mp 1}$, whereas every odd permutation $\tau \in S_3 \setminus A_3$ maps $\ket{W_+}$ to $\ket{W_-}$ up to a phase. Thus, the projective $S_3$-orbit of $\ket{W_+}$ contains precisely two orthogonal codewords, $\ket{W_+}$ and $\ket{W_-}$.

To verify the noiseless-subsystem property, note that both $\ket{W_+}$ and $\ket{W_-}$ belong to the single-excitation sector ($S^z_{\mathrm{tot}} = +1/2$) and are annihilated by the collective spin-raising operator
\begin{equation}
S^+_{\mathrm{tot}} = \sum_{n=1}^3 \sigma_n^+,
\end{equation}
as follows from
\begin{equation}
S^+_{\mathrm{tot}}\ket{W_\pm} = \frac{1}{\sqrt{3}}\left(\omega^{\mp 1} + 1 + \omega^{\pm 1}\right)\ket{000} = 0.
\end{equation}
Hence $\ket{W_+}$ and $\ket{W_-}$ are both highest-weight states of the total spin-$1/2$ ($\bm{S}_{\mathrm{tot}}^2 = \frac{3}{4}$) irreducible representation of $\mathrm{SU}(2)$. Under Schur--Weyl duality,
\begin{equation}\label{eq:SchurWeyl}
(\mathbb{C}^2)^{\otimes 3} \cong \left(\mathcal{H}_{J=3/2} \otimes \mathbb{C}^1\right) \oplus \left(\mathcal{H}_{J=1/2} \otimes \mathbb{C}^2\right),
\end{equation}
the pair $\{\ket{W_+}, \ket{W_-}\}$ spans the two-dimensional multiplicity space $\mathbb{C}^2$ associated with the $S^z_{\mathrm{tot}} = +1/2$ weight of the $J=1/2$ sector. Because every collective Kraus operator commutes with $S_3$ and acts trivially on the multiplicity space,
\begin{equation}\label{eq:SchurKraus}
E_\mu^{\otimes 3} = \bigoplus_{J \in \{1/2, 3/2\}} E_{\mu,J} \otimes I_{\mathrm{mult}},
\end{equation}
any encoded logical state
\begin{equation}\label{eq:logical_state}
\ket{\psi_L} = \alpha\ket{W_+} + \beta\ket{W_-} \in \mathcal{C}_{\mathrm{NS}} \qquad \left(|\alpha|^2 + |\beta|^2 = 1\right)
\end{equation}
evolves under collective errors into
\begin{equation}\label{eq:NS_output}
\mathcal{E}\left(\ket{\psi_L}\bra{\psi_L}\right) = \rho_{J=1/2} \otimes \ket{\psi_L}\bra{\psi_L},
\end{equation}
preserving the encoded logical qubit without active error correction.
\end{proof}

Furthermore, as shown in Eqs.~\eqref{eq:Ep_ring}--\eqref{eq:Em_ring}, tuning a Hermitian three-qubit ring to $2\sqrt{3}D = J_\chi$ with $J > 0$ renders $\mathcal{C}_{\mathrm{NS}}$ an exact twofold-degenerate ground-state manifold protected by an energy gap $6J$, providing passive energetic suppression against excitations out of the single-excitation sector in addition to symmetry protection against collective noise.

\subsection{Implications for controllable $XX$ and $XXZ$ qubit arrays}\label{ControlImplic}
Finally, the symmetry and $\mathrm{LU}$ structure established in Secs.~\ref{sec:circulant}--\ref{sec:LU} directly governs the dynamical preparation of chiral $W$ states in one-dimensional spin-$1/2$ chains. Consider a linear array of three nearest-neighbor $XX$- or $XXZ$-coupled qubits subject to a single local $Z$ control field acting exclusively on the first qubit:
\begin{align}\label{eq:H_chain}
\lefteqn{H_{\mathrm{chain}}(t) = u(t)\sigma_1^z} \nonumber \\
&\quad + J\sum_{n=1}^{2}\left(\sigma_n^x \sigma_{n+1}^x + \sigma_n^y \sigma_{n+1}^y + \Delta \sigma_n^z \sigma_{n+1}^z\right),
\end{align}
where $u(t)$ is a tunable local control amplitude. Because both the drift interaction and the local $\sigma_1^z$ control conserve the total excitation number
\begin{equation}\label{eq:N_op}
\hat{N} = \sum_{n=1}^3 \ket{1}_n\!\bra{1},
\end{equation}
the system cannot be fully controllable on the entire eight-dimensional Hilbert space $(\mathbb{C}^2)^{\otimes 3}$. However, according to Lie-algebraic controllability criteria~\cite{DAlessandro:07}, such arrays are controllable on the three-dimensional single-excitation subspace $\mathcal{H}_1$ for both the noninteracting $XX$ model ($\Delta = 0$) and the interacting $XXZ$ model ($\Delta \neq 0$); moreover, in the interacting $XXZ$ case, the system is controllable on every subspace with a fixed number of excitations.

Three structural features from Secs.~\ref{sec:circulant}--\ref{sec:LU} make this control architecture both necessary and sufficient for synthesizing arbitrary chiral $W$ states:
\begin{enumerate}
\item \emph{Completeness in $\mathcal{H}_1$:} Because $\{\ket{W_3}, \ket{W_+}, \ket{W_-}\}$ spans the entire single-excitation subspace $\mathcal{H}_1$, controllability on $\mathcal{H}_1$ is sufficient to steer a localized excitation $\ket{100}$ into $\ket{W_+}$, $\ket{W_-}$, or any logical superposition $\ket{\psi_L} \in \mathcal{C}_{\mathrm{NS}}$ in Eq.~\eqref{eq:logical_state}.
\item \emph{Necessity of breaking $\mathbb{Z}_3$ symmetry:} By Proposition~\ref{prop:LU}(ii), global single-qubit unitaries $U^{\otimes 3}$ preserve $\mathbb{Z}_3$ characters and cannot induce transitions among $\ket{W_3}$, $\ket{W_+}$, and $\ket{W_-}$. Thus, global control on a homogeneous ring is fundamentally incapable of generating chirality from $\ket{W_3}$; breaking cyclic permutation symmetry via open boundaries and local single-qubit addressing $u(t)\sigma_1^z$ is essential.
\item \emph{Local phase interconversion:} By Proposition~\ref{prop:LU}(i), once any one of the states $\ket{W_3}$, $\ket{W_+}$, or $\ket{W_-}$ is prepared, the remaining two can be reached deterministically via single-qubit diagonal phase gates,
\begin{equation}
D_+ = V^\dagger \otimes I \otimes V, \qquad D_- = V \otimes I \otimes V^\dagger.
\end{equation}
\end{enumerate}

\section{Summary and conclusions}\label{SummConcl}

A unified algebraic, group-theoretic, and entanglement characterization of the chiral three-qubit $W$ states $\ket{W_+}$ and $\ket{W_-}$ has been presented in this work. 

By formulating the single-excitation sector in terms of general $3\times 3$ circulant matrices and the one-dimensional irreducible representations of $\mathbb{Z}_3 \cong A_3$, it has been demonstrated that $\{\ket{W_3}, \ket{W_+}, \ket{W_-}\}$ serves as the universal discrete Fourier eigenbasis for both coherent Hermitian spin rings with Dzyaloshinskii--Moriya and scalar chirality interactions and open non-Hermitian rings with nonreciprocal hopping. 

Specializing general results on $A_3$-invariant states~\cite{Lyons+:22} to the chiral-$W$ manifold, it has been shown that while arbitrary superpositions of $\ket{W_\pm}$ and their spin-flipped two-excitation counterparts $\ket{\overline{W}_\pm}$ share identical bipartite and tripartite entanglement invariants (constant single-qubit purity, equal pairwise concurrences, and vanishing three-tangle), only the pure Hamming-weight-$1$ and Hamming-weight-$2$ basis states are exact eigenstates of the scalar chirality operator. In addition, it has been proved that $\ket{W_+}$, $\ket{W_-}$, and $\ket{W_3}$ share a single $\mathrm{LU}$ equivalence class under arbitrary single-qubit unitaries $\mathrm{U}(2)^{\otimes 3}$ while remaining strictly inequivalent under symmetric global unitaries $U^{\otimes 3}$, exact Bloch-angle criteria have been derived for $A_3$-symmetrization of product states, and their continuous and discrete local stabilizer groups have been characterized. Finally, the orthogonal $S_3$-orbit $\{\ket{W_+}, \ket{W_-}\}$ has been shown to realize a minimal permutation-orbit noiseless subsystem, and the resulting symmetry-breaking requirements governing controllability on the single-excitation subspace in linear $XX$ and $XXZ$ qubit arrays have been established.

\end{document}